\newcounter{myctr}
\def\myitem{\refstepcounter{myctr}\bibfont\noindent\ifnum\themyctr>9\else\phantom{0}\fi\hangindent17pt\themyctr.\enskip}

\documentclass{ws-ijqi}

\begin{document}

\markboth{Suranjana Ghosh} {Coherent control of mesoscopic
superpositions in a diatomic molecule}

\catchline{}{}{}{}{}

\title{COHERENT CONTROL OF MESOSCOPIC SUPERPOSITIONS \\
       IN A DIATOMIC MOLECULE}

\author{SURANJANA GHOSH}

\address{Indian Institute of Technology Patna, Patliputra Colony \\
         Patna 800013, India \\sghosh@iitp.ac.in}

\maketitle

\begin{history}
\received{Day Month Year}
\accepted{Day Month Year}
\end{history}

\begin{abstract}
A phase controlled wave packet, recently used in experiment of
wave packet interferometry of a diatomic molecule, is investigated
to obtain mesoscopic superposition structures, useful in quantum
metrology. This analysis provides a new way of obtaining
sub-Planck scale structures at smaller time scale of revival
dynamics. We study a number of situations for delineating the
smallest interference structures and their control by tailoring
the relative phase between two subsidiary wave packets. We also
find the most appropriate state, so far, for high precision
parameter estimation in a system of diatomic molecule.
\end{abstract}


\section{Introduction}

Mesoscopic superposition in different systems has become a
benchmark in quantum information processing and quantum metrology
\cite{schleich1,schleich2,schleich3,schleich4,davidovich,raimond,auffeves}.
Nonlocal quantum superposition can produce quantum interference
structure of dimension below Planck scale, well known as
sub-Planck structure (SPS). After the invention of SPS by Zurek
\cite{zurek}, this has attracted an enormous physicists' attention
\cite{ph,pathak,toscano,ghosh1,dalvit,praxmeyer,jay,bhatt,milburn,scott,walmsley,jafarov,super}.
The fact that these are very sensitive against any external
perturbation or decoherence makes them useful for high precision
quantum parameter estimation and quantum metrology
\cite{toscano,ghosh1,ghosh2,roy}. Other than systems modeled by
harmonic oscillator, obeying Heisenberg-Weyl algebra, SPS was also
found in various solvable quantum mechanical potentials, having
SU(1,1) and SU(2) symmetries and nonlinear energy spectrums
\cite{ghosh1,ghosh2,roy}. Particularly in diatomic molecule
(modeled by the Morse potential), SPS was first found in the long
time evolution of an appropriately framed coherent state
\cite{ghosh1}. This application recently brought forward towards
the important sensitivity analysis and the effect of decoherence
\cite{ghosh2}. In the present work, we undertake a new approach,
inferred from the experiment where two femtosecond laser pulses
whose relative phase was controlled to create a phase-locked
vibrational wave packets in iodine molecule
\cite{fleming1,fleming2,ohmori10}. Recently, in another remarkable
work, a spatiotemporal images of quantum interference on the
picometer and femtosecond scale has been reported by using their
ultra-precision wave packet interferometry technique
\cite{ohmori09}. This technique has precise control over the
relative phase of the twin optical pulses as well as on the two
component subsidiary wave packets. Thus, this method would be
capable of monitoring the quantum interference ripples of the
phase-locked wave packet. Controlling quantum interference
patterns by various means has enormous importance in quantum
information science. Here, we show the way of controlling the
mesoscopic superposition structures and discuss some results which
can't be achieved in the earlier studies using a single localized
wave packet dynamics \cite{ghosh1,ohmorisc}. We study this phase
locked wave packet, where the relative phase between the two pump
pulses plays a crucial role in the system dynamics. This coupled
wave packet is capable to capture the signature of the sub-Planck
interference structures in Wigner distribution at earlier time as
compared to the single coherent state dynamics. In the next
section, we elaborate and explain the phase-controlled wave
packet, which is followed by the control of the mesoscopic
superpositions in spatial domain as well as in phase space. We
show how by tuning the relative phase of the wave packets one can
actively tailor the quantum interference structures. An systematic
numerical analysis yields a compelling evidence to find the most
appropriate state, so far in literature, for high precision
parameter estimation and quantum metrology in the system of
diatomic molecule. We end with some conclusions and future
outlook.

\section{The phase-control wave packet}
A wave packet, consisting of two partial wave packets, generated
with a pair of femtoseceond laser pulses whose relative phase is
coherently controlled, can be written as \cite{ohmori09}

\begin{equation}\label{wavepacket}
\Phi_{\theta}(\xi,t) = \frac{1}{2}\left[(1-e^{i \theta})
\Phi_{1}(\xi,t)+(1+e^{i \theta}) \Phi_{2}(\xi,t)\right],
\end{equation}
where $\Phi_{1}(\xi,t)$ and $\Phi_{2}(\xi,t)$ are composed of even
and odd vibrational levels of a diatomic molecule modeled by Morse
potential \cite{ghosh1}. $\xi$ is the variable, related to the
internuclear distance, ranging from $0<\xi<\infty$. The parameter
$\theta$ defines the pump-control phase which can be arbitrarily
tuned between $0$ and $2\pi$. As coherent state is most classical
in quantum framework and most suitable for studying wave packet
dynamics \cite{perelomov}, we use an appropriate SU(2) coherent
state, consistent of its dynamical symmetry having finite number
of bound states \cite{ghosh1}. Individually, the states
$\Phi_{1}(\xi,t)$ and $\Phi_{2}(\xi,t)$ can also be constructed
analytically by using a quadratic algebra, which produces
superposition of alternate energy levels. This phase locked wave
packet allows a nice control over the component $\Phi_{1}(\xi,t)$
and $\Phi_{2}(\xi,t)$ by tuning the relative optical phases of the
laser pulses.
\begin{figure*}\centering
\includegraphics[width=3.2in]{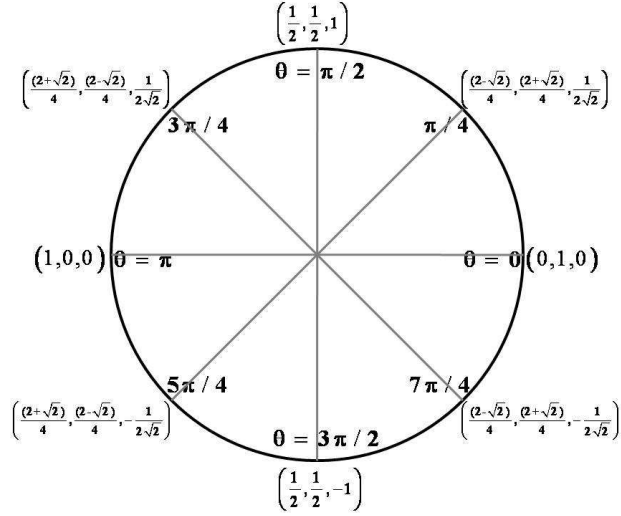}
\caption{Phase-circle showing the probability density of the phase
controlled wave packet for different $\theta$s. Three indices
inside the parentheses show the contribution of the even, odd and
cross-terms modulo $|\Phi|_{1}^2$, $|\Phi|_{2}^2$ and
$i\sqrt{2}(\Phi_{1}^{*}\Phi_{2}-\Phi_{2}^{*}\Phi_{1})$,
respectively.} \label{phase}
\end{figure*}
The even and odd states result at $\theta=\pi$ and $\theta=0$,
respectively. Otherwise, it is a mixture of both the components
involving all energy levels with conserved probability at an
arbitrary time $t$. The probability density at any time is given
by
\begin{eqnarray}\label{prob}
|\Phi_{\theta}(\xi,t)|^2 &=&\frac{1}{4}\left[(1-\cos{\theta})
|\Phi_{1}(\xi,t)|^2+(1+\cos{\theta})
|\Phi_{2}(\xi,t)|^2\right.\nonumber\\
&&+\left.
i\sin{\theta}\left(\Phi_{2}(\xi,t)\Phi_{1}^{*}(\xi,t)-\Phi_{1}(\xi,t)\Phi_{2}^{*}(\xi,t)\right)
\right].
\end{eqnarray}
The phase circle in Fig.~\ref{phase} shows the coefficient of even
part, odd part and the cross term of the probability density for
different values of $\theta$ in step of $\pi/4$. It is apparent
that the cross term for any two states of phase difference `$\pi$'
is opposite, which explains the result found in the study of
quantum carpets \cite{ohmori09}. Now, we define $\Phi_{1}(\xi,t)$
and $\Phi_{2}(\xi,t)$ as follows
\begin{eqnarray}\label{evenodd}
\Phi_{1}(\xi,t)=\sum_{m_{1}=0}^{m'}d_{m_{1}}
\;\psi_{m_{1}}(\xi) \exp^{-iE_{m_{1}}t}\nonumber\\
\Phi_{2}(\xi,t)=\sum_{m_{2}=0}^{m'}d_{m_{2}} \;\psi_{m_{2}}(\xi)
\exp^{-iE_{m_{2}}t}
\end{eqnarray}
where $m_{1}$ and $m_{2}$ stand for vibrational quantum numbers
corresponding to the even and odd eigenstates $\psi_{m_{1}}(\xi)$
and $\psi_{m_{2}}(\xi)$, respectively.  Eigenvalues are given by $
E_m=-(D/\lambda^{2})(\lambda-m-1/2)^2$ defining the classical and
the revival times, $T_{\mathrm{cl}}=T_{\mathrm{rev}}/(2
\lambda-1)$ and $T_{\mathrm{rev}}=2\pi\lambda^{2}/D$,
respectively. $d_{m}$ yields the weighting factor of the SU(2)
coherent state. A detail description and full expressions are
given in Ref. \cite{ghosh1}. We have used atomic unit ($\hbar=1$)
throughout our study. Here, we consider $I_2$ molecule and the
corresponding parameter values: potential parameter $\beta=4.954$,
reduced mass $\mu=11.56\times 10^4$, equilibrium inter-nuclear
distance $r_0=5.03$, and dissociation energy $D=0.057$. Coherent
state parameter is taken as $\alpha=2$, which includes $24$ lower
bound states of this molecule.

\section{Results and Discussions: Control of mesoscopic
superpositions and sensitivity analysis}

Initially, even and odd probability amplitudes are same and the
cross-terms do not exist. Thus, the initial wave packet itself is
a combination of two localized components and behaves like a
Schr\"{o}dinger-cat state for arbitrary $\theta$. Now it is
interesting to see what will happen for longer time evolution. On
the other hand, one can observe the changes in the interference
structures due to controlled mesoscopic superpositions at
particular time. In our further study we mainly concentrate on two
typical fractional revival times
\cite{parker,alber,averbukh,robinett} $\frac{1}{8}T_{rev}$ and
$\frac{1}{16}T_{rev}$. The reason behind this are: First,
single-wave packet dynamics (SWPD) at $t=\frac{1}{8}T_{rev}$
results the well studied sub-Planck scale structures originated
from a compass-like state (a four way split of a coherent state
situated in north-south-east-west directions) in phase space.
Hence, it is worth to study what happens by tailoring the phase of
the presently considered wave packet. Second, at
$t=\frac{1}{16}T_{rev}$, SWPD results a eight-fold wave packet or
a combination of two imprinted compass-like states. Thus, this
state would be more interesting to investigate in the present
scenario. Moreover, the coherent control admits both lower and
higher order mesoscopic superpositions at a particular time in
contrast to the SWPD. So we expect some interesting features in
both the cases, which are the consequences of cross-diagonal
superpositions. It is difficult to resolve further higher order
superpositions at higher order fractional revival times due to
limited phase space support and asymmetry of the potential.
\begin{figure}[htpb]
\centering
\includegraphics[width=6.4 in]{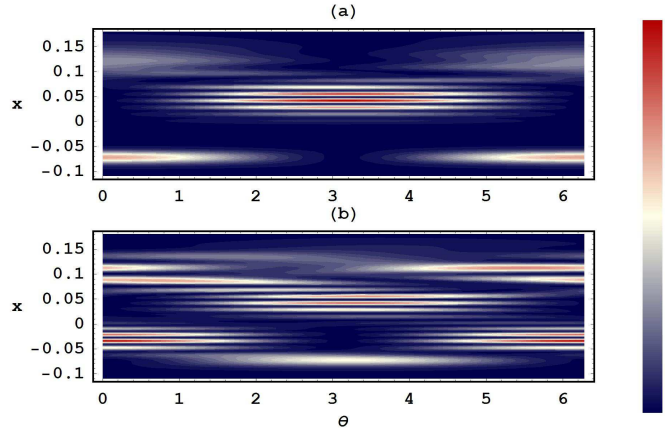}
\caption{(Color online) Quantum carpet shows the phase-control of
the probability density at one-eight (a) and one-sixteen (b) of
revival times, respectively. Here $x=r/r_{0}-1$, is the
dimensionless variable, $r$ and $r_{0}$ both are in the a.u,
whereas $\theta$ varies between $0\leq\theta\leq2\pi$ radians.}
\label{cont}
\end{figure}
\subsection{Control of Spatial Ripples}
We start with the control of spatial ripples, through the
exploration of quantum carpets tailored by $\theta$ at two
specific times mentioned above (see Fig.~\ref{cont}). The quantum
carpet is characterized by the maxima and minima of probability
stretching out in a space-phase representation. At
$t=\frac{1}{8}T_{rev}$, Fig.~\ref{cont}(a) shows how the
controlled relative phase designs the present carpet. For
$\theta=0$, probability distribution resembles the initial
phase-locked wave packet, which is a cat-state. However, the
spatial interference ripples begin to rise and acquire maximum
value at $\theta=\pi$, where the component wave packets fully
overlap in coordinate space. The ripples vanishes again at
$\theta=2\pi$, resulting a spatially separated cat state. The
second case, $t=\frac{1}{16}T_{rev}$, as shown in
Figure~\ref{cont}(b), is bit nontrivial.
\begin{table}[h]
\tbl{Variation of the amplitude of interference maxima $A_{m}$
with $\theta$ at $t=\frac{1}{8}T_{rev}$. $A_{m}$ has dimension
$[L]^{-1}$ and is in atomic unit.}
{\begin{tabular}{|l||l|l|l|l|l|l|l|l|l|} \hline $\theta$ & $ 0 $ &
$\pi/8$  &  $\pi/4$  &  $3\pi/8$  & $\pi/2$ & $5\pi/8$ & $3\pi/4$
&  $7\pi/8$  &  $\pi$  \\ \hline $A_{m}$ & $0.0$ & $ 0.305$ &
$1.14$ &
$2.39$ & $3.85$ & $5.31$ & $6.54$ & $7.35$ & $7.63$ \\
\hline
\end{tabular}}
\label{table}
\end{table}
Spatial ripples are present both at initial and final stages. In
addition, we get some extra ripples in between, interpretation of
which is not obvious from this plot. It captures mostly the
superposition structures instead of individual counterparts. For
further clarification, we will recall it while explaining the
phase space distribution. In the first case, we have numerically
estimated the amplitude of the highest interference ripples
($A_{m}$) at $\frac{1}{8}T_{rev}$ (Table.~\ref{table}). It reveals
the influence of $\theta$ on $A_{m}$, where $A_{m}$ starts from
$0$ at $\theta=0$ and gradually regains maximum value for
$\theta=\pi$. For the rest of the interval, $\pi$ to $2\pi$, it is
a mirror image of the first half. It is worthy to note that for a
fixed evolution time, one is able to control quantum ripples which
signifies the signature of quantum character.

\subsection{Control of Phase-space structures}
The mesoscopic superposition structures are the result of nonlocal
superpositions of quantum states, best displayed in the phase
space Wigner representation:
\begin{equation}\label{wigner1}
W(x,p,t)=\frac{r_{0}}{\pi}\int_{-\infty}^{+\infty}
\Phi^{*}(x-x',t)\Phi(x+x',t) e^{-2ipx'}dx'\;,
\end{equation}
where $x$ coordinate is related with $\xi$ as $\xi=2\lambda
e^{-\beta x}$ and $x=r/r_{0}-1$. Negative regions in the
oscillatory structures of this function indicate non-
classicality.
\begin{figure}[htpb]
\includegraphics[width=5.4 in]{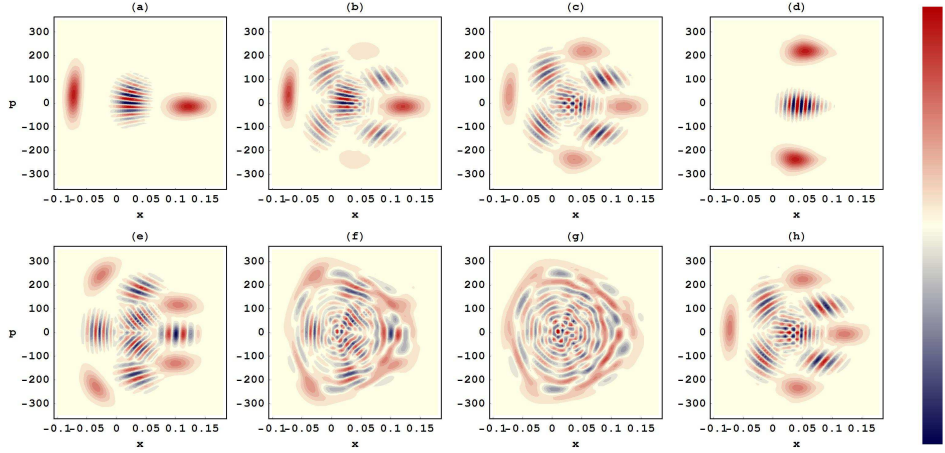}
\caption{(Color online) Wigner distribution of phase locked wave
packets at $t=\frac{1}{8}T_{rev}$ (a)-(d) and
$t=\frac{1}{16}T_{rev}$ (e)-(h) for different $\theta$ ($0$,
$\pi/4$, $\pi/2$, and $\pi$). Here, $x$ and $p$ are the
dimensionless position and momentum, where $x=r/r_{0}-1$ and $p$
is the corresponding scaled variable.} \label{wigall}
\end{figure}
Sub-Planck scale structures usually appear as alternate small
`tiles' of maxima and minima. The phase space area of these
`tiles' are much less than $\hbar$ ($1$ in a.u.). It is inversely
proportional to the effective phase-space area globally occupied
by the state, which can be significantly larger than $\hbar$.
These structures are very sensitive to environmental decoherence
\cite{zurek,pathak,ghosh2} and have important applications in
Heisenberg-limited measurements and quantum parameter estimation
\cite{toscano,dalvit,roy}. We have shown schematically the Wigner
functions at $t=\frac{1}{8}T_{rev}$ (Fig.~\ref{wigall} (a)-(d))
and $t=\frac{1}{16}T_{rev}$ (Fig.~\ref{wigall} (e)-(h)) for
different $\theta$s. We show how these structures can be
controlled by tailoring the phase. In Fig.~\ref{wigall}(a), null
phase signifies a cat state thus the initial phase locked wave
packet is reviving at $\frac{1}{8}T_{rev}$. This is significantly
different from SWPD. The exact contribution of even and odd parts
of probability density can be seen from the phase circle (see
Fig.~\ref{phase}). Another crucial factor is how even and odd
states behave at some particular time. At certain time, the
diversity in mesoscopic superposition structures arise due to the
competing two components associated with even and odd parts. In
Fig.~\ref{wigall}, first row manifests compass-like state for all
$\theta$ except $\theta=0$ and $\pi$. In the last case
(Fig.~\ref{wigall}(d)), this is again a cat state localized in
momentum regime. Thus, one can monitor the interference ripples,
either in position or in momentum space. With increasing value of
$\theta$, even parts starts to contribute appreciably (see
Fig.~\ref{wigall}(b)). At $\theta=\pi/2$, we obtain well define
sub-Planck tiles originated from compass-like state where the even
and odd parts contributed commensurately. The structures at
$t=\frac{1}{16}T_{rev}$ (see second row of Fig.~\ref{wigall}) are
more interesting. In Fig.~\ref{wigall}(e) and \ref{wigall}(h), the
exact control over $\theta$ gives rise to compass-like states at
earlier time than SWPD  in which it usually takes place at
$\frac{1}{8}T_{rev}$. Moreover, \ref{wigall}(e) reveals a
different kind of compass state (not a
`north-south-east-west'-kind like \ref{wigall}(h) and
\ref{wigall}(c)), may have extra importance as we are dealing with
the asymmetric potential. Figure \ref{wigall}(f) and
\ref{wigall}(g) both are eight-fold mesoscopic superposition or
overlap of two compass-like states. A better explanation of the
quantum carpet as shown in Fig.~\ref{cont}(b) can be found from
the phase space description as displayed in
Fig.~\ref{wigall}(e)-(h). We need to perform a quantitative
analysis for revealing the advantages of this method than the
earlier one \cite{ghosh1}.

\subsection{Quantitative study}
The smallest structures in the interference regime appear as
alternate positive and negative tiles of sub-Planck dimension. The
phase-space area occupied by the tiles `$a$' scales as
$\sim\hbar^2/A$ ($\sim 1/A$ in atomic units), where $A$ is the
classical action of the state.
\begin{table}[h]
\tbl{Tailoring the dimension (in atomic unit) of sensitive
mesoscopic superposition (sub-Planck) structures with $\theta$ at
two typical times.} {\begin{tabular}{|l||l|l|l|l|l|l|l|l|l|}
\hline  $\;\;\theta$ & $0$ & $\pi/8$ & $\pi/4$ & $3\pi/8$ & $\pi/2$ & $5\pi/8$ & $3\pi/4$ & $7\pi/8$ & $\pi$ \\
\hline\hline $\frac{1}{8}T_{rev}$ & $0.185$ & $0.148$ & $0.108$ & $0.089$ & $0.083$ & $0.089$ & $0.109$ & $0.157$ & $0.210$ \\
\hline
$\frac{1}{16}T_{rev}$ & $0.0766$ & $0.0769$ & $0.0776$ & $0.0786$ & $0.0800$ & $0.0814$ & $0.0827$ & $0.0835$ & $0.0837$\\
\hline
\end{tabular}}
\label{table2}
\end{table}
$A$ is approximately given by the product of the effective support
of its state in position and momentum: $A\sim \Delta x\times
\Delta p$, where $\Delta x=\sqrt{\langle x^2 \rangle - \langle x
\rangle^2}$ and $\Delta p=\sqrt{\langle p^2 \rangle - \langle p
\rangle^2}$. In Table~\ref{table2}, we have numerically evaluated
the uncertainty products and the area of the smallest tiles both
at $\frac{1}{8}T_{rev}$ (second row) and $\frac{1}{16}T_{rev}$
(third row), respectively. The most important property of these
sub-Planck tiles is the fact that the corresponding quantum state
is extremely sensitive to external perturbation. This sensitivity
is also represented by the well-known formula \cite{schleich2}:
\begin{equation}
|\langle\Phi|\Phi'\rangle|^2=\int\int W(x,p) W'(x,p) dx\; dp,
\end{equation}
where $W$, $W'$ are the Wigner functions of the unperturbed
$|\Phi\rangle$ and the perturbed $|\Phi' \rangle$ quantum states
respectively. The perturbed and unperturbed states become
distinguishable when a single tile (maxima or minima) of the
interference structure is displaced due to a small external
perturbation and its maximum coincides with a minimum of the
undisplaced state. Thus, one has destructive interference and the
two states become approximately orthogonal and distinguishable.
Therefore these structures determine the limit of sensitivity of
the system. We observe few robust behaviors: first, if we compare
column-wise, area of the tiles in the third row is always smaller
than that of the second row for all $\theta$. Hence, the
structures in Fig.~\ref{wigall} (e)-(h) are more sensitive than in
Fig.~\ref{wigall} (a)-(d); second, the tiles in the three cases,
{\it i.e.}, in Fig.~\ref{wigall}(c), Fig.~\ref{wigall}(e), and
Fig.~\ref{wigall}(h), are all originated from compass-like states.
Upon comparing the area, we find that the structures of the new
compass-like state in Fig.~\ref{wigall}(e) are smallest and most
sensitive among these three cases. In fact, this provides the
smallest structures among all the other mesoscopic superpositions
structures in phase space hence improving the sensitivity limit in
the literature.

\subsection{Conclusions}

We have shown a way to control the sensitive mesoscopic
superposition structures in a diatomic molecule. The wave packet
can be observed at a particular time and then relative phase
between the two pulses can be controlled coherently. The
sub-Planck interference structures play very important role in
high precision parameter estimation and quantum metrology. Even at
a definite time, we show how one can obtain either higher or lower
order superpositions by tailoring the phase. It is worth
mentioning that one of the difficulties to realize these
structures in experiments is the default system-environment
interaction, which results the structures decay before reaching to
that desired time. Although, very recently there are few
experiments demonstrating decoherence in a controlled way, this
issue still remains a big hurdle in quantum information science.
Hence, physicists always keep their eyes to find various methods
to create SPS in quantum systems. Primarily, we show the way to
create and control these structures in diatomic molecule. In
addition, it has also found several extra advantages over the
earlier one. First of all, our method is capable of obtaining the
SPS at smaller evolving time with respect to earlier methods and
this minimizes the difficulty to observe these structures in
presence of decoherence. Secondly, we indicate the most sensitive
(smallest) structures, appeared in a new type of compass state in
diatomic molecule. We intend to study the decoherence of this
state in future, and also look over the application in quantum
computation.

\section{Acknowledgments}
The author acknowledges the financial support provided by the
Department of Science and Technology, Govt. of India (Fast Track
project No. $SR/FTP/PS-062/2010)$.

\end{document}